# Design and evaluation of a multi-sensory representation of scientific data

**Stefania Varano** [1, *] **and Anita Zanella** [2]

[1] *Istituto Nazionale di Astrofisica, Istituto di Radioastronomia, via Gobetti 101, 40129 Bologna, Italy*
[2] *Istituto Nazionale di Astrofisica, Osservatorio di Padova, Vicolo dell'Osservatorio 5, 35122 Padova, Italy*

Correspondence*:
Stefania Varano
stefania.varano@inaf.it

## ABSTRACT

**Introduction:** Modern sciences and astrophysics in particular study objects and phenomena not visible in physical terms, that is they cannot be investigated with the eyes or analogous optical systems. Nevertheless, they make intensive use of visual representations, showing data in a figurative way, using lights, shadows, colors, and shapes familiar to the user and aesthetically pleasant. Besides being inaccessible for Blind and Visually Impaired (BVI) users, such figurative visual representation can lead to misunderstandings about the real nature of the represented object, particularly if the representation code is not declared. In this work we argue that multi-sensory representations clearly arbitrary, not aiming to imitate reality, are a valid choice for an effective and equal engagement with astronomy.

**Methods:** To investigate our hypothesis, we designed two mono-sensory representations (haptic only and acoustic only) and tested them individually and in group workshops with both sighted and BVI users, running a qualitative assessment about the usability, understandability and non-ambiguity of the representations. During the experimentation we collected recommendations about most suitable modes of representation and mapping of represented features. These findings were used to guide the design of a multi-sensory representation of non-visible astronomical data including visual, acoustic, and haptic stimuli. We tested also this representation, to refine the product and propose it to the public.

**Results:** The outcome of our work is the exhibit "Sense the Universe", which was part of a museum exhibition attended both by sighted and BVI users. The responses referred to the representation as pleasant, easily usable by all and non-ambiguous, namely not leading to misinterpretations related to earlier knowledge and experience, indicating that the iterative design procedure was effective for the purposes we aimed at. The presence of a cognitively challenging code of representation made the fruition conscious and attentive. In an equity perspective, Sense the Universe proven to create a true common ground among people with diverse abilities, skills, and learning styles, in the framework of Universal Design for Learning.

**Discussion:** Our findings suggest the validity of multi-sensory representations for a true and effective engagement with science, its contents and processes and for providing an equal access to scientific culture, both in outreach and education contexts.

# 1  INTRODUCTION

"A picture is worth a thousand words", says a very ancient aphorism. It is commonly believed that visual representations are the most effective form of communication. In science they have often been fundamental for building the perception of scientific results, not only in terms of understanding a concept, but also as a powerful imaginative impulse (Barrow, 2009). Visual representation of scientific data are usually divided in "information" and "scientific" visualizations (Card et al., 1999; Friendly, 2008; Munzner, 2008). The difference between the two is in the nature of the represented data: scientific visualization deals with physical based data, representing concrete quantities, while information visualization deals with more abstract entities. Examples of use of the two in astrophysics are the visualizations of the temperature or emission intensity of an object (scientific visualization) and on the other hand the classification of groups of objects or their statistical distribution (information visualization). Scientific visualizations in astrophysics are often *figurative*, meaning that they are tied to the spatial geometry and morphology of the represented object. They can have greater or lesser *figurative density* (Greimas, 1984), from representations that strongly imitate reality (e.g., images that make use of realistic colours for representing regions at different intensity or different shells of the inner parts of the Sun) up to completely arbitrary representations that use graphic features clearly unrelated to visual experience (contour maps of the intensity emission of a celestial object or lines connecting the magnetic poles of Jupiter). In astrophysics, extensive use of highly figurative scientific visualizations is made; also for astronomical data not visible to the human eye, resulting from direct observations and measurements. An example of this is the digital imaging technique of *false colors*, widely used in astrophysics, that use fictitious scales of colors to represent electromagnetic waves invisible to the human eye (or analogous optical systems, such as optical telescopes), because they do not emit or reflect visible electromagnetic waves, or cannot be hit by visible light (which they would otherwise reflect), or their emitted/reflected light cannot reach the observer. This technique has recently been used to represent the emission of radio waves from the black holes in galaxy M87 and in the Milky Way, obtained by the Event Horizon Telescope collaboration (EHT[1], (Doeleman et al., 2009)) and released in 2019 and 2022, respectively.

In semiotic terms, any visual representation is a *sign*, i.e. something that stands for something else (its *object*) (Peirce, 1931). The correspondence between sign and object is given by the code (Jakobson, 1963) of the representation, i.e. a precise set of interpretative rules.

When scientific figurative visualization of invisible objects and phenomena is used, the user or learner can mistake the (unknown) visual nature of the object with its visual representation.

This applies to any figurative representation, also the ones that make use of other sensory stimuli (acoustic, haptic), that are strictly connected to possible sensory experience and can make the real nature of the object misunderstood.

To avoid such misconceptions, we argue that highly figurative representations should be always accompanied by a declaration of their disconnection from reality (e.g. "false colors" or "artist impression", in the case of visuals) and possibly by the explication of the code of the representation.

---

[1] EHT collaboration link

This applies both to academic contexts (peer to peer communication among professionals) and to public outreach (in mass media communication, events, science museums, and so on), and is unavoidable in educational contexts. On the one hand because the meaning-making process requires a true engagement with the representation, but also because the process that leads to the production of any scientific representation is part of its beauty and power, and should be properly addressed, in the framework of science process skills teaching and learning (Pantidos et al., 2022; McComas, 2014).

## 1.1 Towards authentic, universal and significant design processes

The ambitious ideal of this study is to contribute to restore the objectivity of scientific data, i.e. numerical records produced by any instrumental measurement. In order to avoid misconceptions and effectively convey the scientific contents and method, we argue that the use of **arbitrary multi-sensory representations** in public engagement with astrophysics and sciences in general is a valid choice for an effective learning and understanding of science. The use of arbitrary (non-figurative), redundant (in which the data is presented through different sensory channels) and multi-modal (in which access to data can occur with different types of interaction) sensory representations can be a valid choice in order to obtain a result as accurate and significant as possible. Such representations, unlikely to be interpreted as an imitation of reality, naturally suggest the presence of a code of interpretation and can help the conceptual understanding (Olympiou et al., 2013).

In an equity perspective, in the framework of Universal Design for Learning (Rose and Meyer, 2002), namely the design of resources for an *equitable use*, without need for arrangements, arbitrary multi-sensory representations can be very effective to create an effective common ground for inclusion among people with diverse abilities and learning styles.

It is impossible to design environments, experiences and activities that can be used in a meaningful way by all (*barrier-free utopia*, (Shakespeare et al., 2006)). Everyone has a different learning style, different life experiences, skills, interests, abilities. The Universal Design for Learning guidelines (CAST, 2018) define the necessity of offering multiple means of engagement, representation and action/expression, that can be summarized as: to offer to the users various ways to find and pursue their own significant meaning in what is presented to them, to grant the users with access to contents and information through different (sensory and stylistic) channels and to give them the opportunity to interact with the content in different modes in order to express themselves.

In this study:

- we investigate existing representations of scientific data that make use of non-visual sensory stimuli, both figurative and arbitrary;
- we design two mono-sensory representations, one making use of sound only and the other with haptic stimuli, and test them separately;
- we use the results of this experimentation to design and evaluate a multi-sensory exhibit representing non-visible astronomical data including visual, acoustic and haptic elements.
- we test the multi-sensory representation as well, and use the outcomes to refine it in order to be presented as an exhibit or a teaching tool in outreach and educational contexts.

The exhibit resulting from this work (named "Sense the Universe") was presented in a public exhibition, where we collected some feedback from both sighted and BVI final users.

## 2 BRIEF LITERATURE REVIEW OF NON-VISUAL REPRESENTATIONS

Understanding how to represent scientific data involving other senses than the sight is an active field of research. Representations alternative to the visual ones include the use of haptic and sound elements. Only ten years ago, alternative scientific representations were rare and were often targeting users with a visual impairment (Harrison et al., 2021; Zanella et al., 2022). Such representations aimed at reproducing an exact correspondence between the image and the sensory map, to compensate for the absence of sight. The number of alternative representations has been constantly growing through time, and now there are dozens of tools available for haptic and sound representations (see the Data Sonification Archive for a constantly updated repository, Lenzi et al. (2020))[2]. The goal of such non-visual representations is diverse: they do not only target blind and visually impaired users, but many of them have specifically been designed to ease the understanding of contents that appear to be more effectively conveyed by the use of touch and/or hearing (Diaz-Merced, 2013).

Despite the large variety of tools that are available nowadays and the efforts of numerous research groups working all over the world, peer-reviewed publications reporting systematic user testing and proper evaluation of the efficacy of non-visual representations is still lacking. Some of the main reasons for this are the relatively recent development of this research field and the multidisciplinary nature of the topic that touches upon astrophysics, social and educational science, psychology, sound design, and computer science (Zanella et al., 2022).

In the following we place our work into context, by mentioning the non-visual representations that were available at the time this experiment was conceived and briefly expand on the multi-sensory representations available nowadays.

### 2.1 Tactile representations

Tactile representations are extensively used to make astronomy accessible to the blind and visually impaired. Ample literature about the fruition process both from a psychological and pedagogic perspective is available (e.g., Fleming (1990); Perkins (2002); Fernandes (2012); Grassini (2015); Horton et al. (2017)).

Tactile and visual perceptions have very different characteristics. Creating a tactile representation does not simply mean to create an image in relief, because the exploration and learning modes in the visual and haptic fruition are different (Claudet, 2009). In the first place, the tactile perception is analytic and sequential, while the visual one is synthetic and instantaneous: vision allows exploring distant phenomena included in the user's field of view, whereas touch requires the contact with the object of investigation and the exploration is limited to the touched area (Iachini and Ruggiero (2010), Ruotolo et al. (2012)). Therefore a tactile representation cannot be the "translation" of the visual one, but it has to be properly designed to convey a manageable amount of information and details (Cattaneo et al., 2008). On the other hand, although the connection between sensory memory and long-term memory is still an active field of research (e.g., Cowan (2009); Roediger and DeSoto (2015), and references therein), sight and touch seem to have in common a strong connection with sensory memory and daily-life experience. This can distort the perception of the users and lead them to misinterpret the phenomenon they are exploring. A complete review of the available tactile representation of astronomical data goes beyond the scope of this paper. We mention here only some examples. The first example we are aware of is a series of books about astronomy: "Touch the stars" (Grice, 2002a), "Touch the Universe – A NASA Braille Book of Astronomy" (Grice, 2002b), "Touch the Sun – A NASA Braille Book" (Grice, 2005b), "The Little Moon Phase Book" (Grice, 2005a), and "Touch the Invisible Sky" (Steel et al., 2008). All these books have been designed to support educators and professors in teaching astronomy to blind students in primary and secondary schools. The last publication in particular contains tactile images created starting from the visualisations of data taken with telescopes operating at different wavelengths (from radio to gamma rays) and available on NASA's archives. It is unclear what the result would have been if the tactile images were created starting from the numerical data directly, before they were turned in visuals. A different approach has been taken by Rule (2011) who created *ad hoc* haptic materials to explain geophysical concepts concerning Mars soil to middle-school students. They have compared the different approach of sighted and visually impaired students to the tactile exploration of the educational material and found that blind users were more "shy" and tended to wait for instructions before exploring the tactile representations. Quantitative and systematic tests on the effectiveness of these representations in conveying the scientific content have not been performed though.

---

[2] Data Sonification Archive: link

More recent non-visual astronomical representations that include haptic elements (in combination with sound and visual elements at times) are: A touch of the Universe[3] and Tactile Universe[4] that provide haptic resources to engage blind and visually impaired students people with astronomy. A4BD[5] that uses vibration to represent the shape of objects; A Dark Tour of the Universe[6] that provide the users three-dimensional prints and tactile models; AstreOS[7], Eclipse Soundscapes[8], and A Universe of Sound[9] which are currently working on including haptic elements in their non-visual representations. The goal of these projects is multi-fold: some are outreach programs targeting the general public, some develop educational material for school pupils, while others are being developed by and for (blind or visually impaired) researchers.

## 2.2 Sound representations

Auditory representations are in most cases *arbitrary*. In most cases the auditory representation is created *ad hoc* starting from the data, rather than from the visuals. Here we present a methodology rather than single representations: the *sonification*, namely the process of turning scientific data into sound. At the time this work started, in 2015, two major physics discoveries were brought to the public attention not only through visual, but also through auditory representations: the discovery of the Higgs boson (July 2012, Aad et al. (2012), Chatrchyan et al. (2012)) and of gravitational waves (September 2015, Abbott et al. (2016)). In the first case, the data taken at the Conseil Européen pour la Recherche Nucléaire (CERN) showing increasing and decreasing trends were sonified by using sounds with increasing or decreasing pitches respectively (link). The arbitrary choice in such representation is the mapping strategy used to sonify the data: while the interval between the data points in the visual representation is given by the sampling of the measurements, the interval between two notes in the auditory representation is set by the chromatic scale (tone, tone, semi-tone, tone, ...). The visual peak revealing the detection of the Higgs boson corresponds to an ensemble of high pitch notes in the auditory representation (Fryer, 2015). The second example we report here is the auditory representation of the gravitational waves detected by LIGO in 2015. In this case, the mapping of data into sound was performed with two consecutive sounds: in the first the frequency of the sound corresponds to the frequency of the detected gravitational wave, while in the second higher frequencies have been used so that they could more easily be perceived by the human ear. During the press conference that followed the discovery of gravitational waves it was said that "you can actually *hear* them, (...) so it's the first time the Universe has *spoken* to us through gravitational waves (...) up to now we were deaf to gravitational waves but now we're able to hear them" (link). The metaphor that sound reveals gravitational waves is powerful and has been used many times afterwards. However it is important to highlight the arbitrariness of this analogy with reality to avoid misconceptions (i.e. we can not directly *hear* the sound of gravitational waves as sound does not propagate in vacuum).

---

[3] A touch of the Universe https://www.uv.es/astrokit/

[4] Tactile Universe https://tactileuniverse.org/

[5] A4BD project: link

[6] A Dark Tour of the Universe: link

[7] AstreOS: link

[8] Eclipse Soundscapes: link

[9] A Universe of Sound: link

Up to 2016 only a handful sonifications of astronomical data were available, whereas in most recent years the number of sonification projects related to astronomy and space science has been growing exponentially (Zanella et al., 2022). A complete review of current attempts to represent astronomical data with sound goes beyond the scope of this paper and we refer the reader to Zanella et al. (2022) for a complete discussion. We notice however that, despite the growing number of sonification projects that are emerging, the documented, peer-reviewed ones are still the minority. Furthermore most of these projects are still lacking proper evaluation. With our study, we aim at setting the stage for future rigorous testing of non-visual data representations and at encouraging the systematic publication of such efforts.

## 3 METHODS

The aim of our study is to develop an arbitrary (non-figurative), multi-sensory exhibit of astronomical data, physically invisible, and to assess the accessibility and intelligibility of such exhibit, in order to use it in schools, museums and public events, even with blind and visually impaired users.

In order to study the suitability of diverse sensory stimuli, we designed two mono-sensory representations, one making use of acoustic stimuli and the other with haptic stimuli, and tested them separately. The experimentation was held in the framework of a research project in collaboration with the "Institute for Blinds" Francesco Cavazza and the Physics Department of Bologna University. We used the results of this experimentation to design a multi-sensory exhibit, representing non-visible astronomical data with arbitrary visual, acoustic and haptic elements. The first version of the multi-sensory representation was evaluated as well, in order to make the needed adjustments in view of the final exhibit. The final exhibit was shown as part of the "Inspiring Stars" exhibition in many events, the last of which was the "Punti di Vista" festival[10], at "Palazzo delle Esposizioni" in Rome, in January 2022. Qualitative feedback has been collected in that occasion as well.

We carried out our experimentation in individual and group workshops, with both sighted and blind or visually impaired users, through an iterative process with the users/learners, in analogy with the design-based learning methodology (Tay, 2004; Barab and Squire, 2004; Collins et al., 2004; Gomez Puente et al., 2013).

For each preliminary experimentation phase, we performed a qualitative evaluation through a questionnaire and a final debriefing, for assessing the validity (and possible limits) of each representation. A qualitative assessment of the final exhibit was also performed, by brief interviews with 2 BVI professional astronomers and some of the people attending the festival "Punti di vista" at "Palazzo delle Esposizioni" where the exhibit was presented.

---

[10] Punti di vista: link

## 3.1 Selection of involved senses

When designing our representations, first we have considered what senses to involve. We have excluded taste and smell for logistical issues and due to existing regulations regarding the delivery of such stimuli. We excluded these senses also for physiological reasons related to the lower flexibility (greater persistence of the sensory stimulus) and the greater subjectivity of these two senses, implying fewer degrees of freedom in the representation (Cain and Potts 1996, Miranda 2012).

Regarding the sense of touch, we did not use the thermal tactile because of limitations in terms of represented values, as touching objects which are too hot ($\gtrsim 40$ degrees Celsius) or too cold ($\lesssim 0$ degrees Celsius) would damage the fingertips of the user. Additionally, it is difficult to recognize too little temperature variations (gradients), further restricting the number of available intervals for the representation (e.g., Bai 2021).

We therefore chose to design sensory representations that make use of sight, touch, and hearing. In order to best exploit such stimuli, we analysed their limits and potentialities, relying on literature data and advice from experts and BVI users.

In particular, we focused on studying both the validity and usability of the sensory stimuli and the modality of interaction of the users with such stimuli. For each stimulus, we considered:

- the informative content: how many diverse characteristics of the data can be included in a representation?
- the perception: how dense can the sampling be in order to still have discernible differences between the different sensory stimuli? For example, in false colours visual representations, how many recognizable nuances of colours can be used and how similar to each other can they be?
- the level of arbitrariness: what is the adequate extent of arbitrariness that can be used, in order to make the symbolic character of the representation still clear and recognizable? How many prompted associations the user must make to effectively relate the representation with the data and their meaning?

A recap of this analysis for visual, haptic and acoustic stimuli is reported in Table 1.

For our design, we used the visual, haptic and acoustic sensory features identified as the most effective and efficient to create a representation rich but still understandable by both BVI and sighted users with diverse ages and scientific backgrounds.

## 3.2 Testing users

We evaluated our exhibit in two phases, with two different samples of users. In the first iterative and pilot phase, we engaged a group of 27 users (see Section 4). We then conducted a qualitative evaluation of the final exhibit with a different sample of 22 users: 2 professional BVI astronomers and 20 visitors of the exhibition "Inspiring Stars" at "Palazzo delle Esposizioni" in Rome, where the exhibit was exposed in January 2022. This second evaluation was intended to validate the results of the main evaluation iterative process.

We chose the first sample of 27 users considering three parameters: age, sight, and scientific literacy. None of the users had previously experienced multi-sensory representations of data (be scientific or not), the sighted users only being familiar with visual representations and the BVI users with haptic ones. The main characteristics of this sample are:

**Table 1.** Analysis of different sensory stimuli in terms of informative content, perception and level of arbitrariness.

| Sensory stimulus | Informative content | Perception | Level of arbitrariness |
| --- | --- | --- | --- |
| **Visual** Digital rendering of images in false colors | Different colours can correspond to different values of a specific feature. | Visual representation is perceived thanks to the chromatic perception of the human eye (three-chromatic and broadband perception of cones). The most effective perception is obtained when colors are intense and very different from each other (Ware, 2013). | Visual representations in false colors showing a varying parameter are straightforward: the use of realistic and natural colors may lead to misinterpreting the representation with the picture of a physical object or system. |
| **Haptic** Relief printing techniques. | Relief points can represent point-like sources. Different haptic features can be used to represent different values of a physical parameter. | There is a minimum height of perceptible relief (the standard for Braille dots is ~ 0.5 mm). *Shadows* and *overlapping* should be avoided. Perspective is not easily understandable by congenitally blind people and scale representations must be trained (Thinus-Blanc and Gaunet, 1997). | It is important to always use different dimension of a tactile feature to represent different physical sizes, in order to avoid misunderstandings (Horton et al., 2017). |
| **Acoustic** Sonification of one-, two-, and three-dimensional data sets. | Sound characteristics (pitch, timbre, volume) can be used to represent individual physical parameters. | The human ear is able to recognize equivalent frequency intervals even at different reference heights. The increase in sound intensity, which is linear, is perceived by the ear in a logarithmic way (Davidovits, 2019). The frequency of a sound is due to a vibration of the medium. The ability to recognize the characteristics of sound depends on acoustic education and musical training (Ediyanto and Kawai, 2019) | Different characteristics of sound can be related to different properties of the represented object. It is important to establish non-ambiguous connections between the characteristics of the sound and the represented features, to avoid misunderstandings and unforeseen connections with memory and experience (e.g. the intensity of sound can easily be connected to the amplitude of the feature, Cowan (2009), Roediger and DeSoto (2015)). |

- **Age**
  All users are adults (aged between 22 and 60), to avoid the complexity of evaluating sensory education and psycho-cognitive training of children, especially in the case of visual impairments.

- **Sight**
  We have involved both sighted (13) and BVI users (14), to pinpoint possible differences in the fruition and understanding of the representation. In fact, we aimed to test senses that for the blind are *vicariant* (i.e. somehow substitute) of sight. We therefore needed to define the effect that a greater training of those senses could have had on the informative power of the representation. On the other hand, we wanted to check if the synesthetic fruition of the visual representation, aside of the haptic and acoustic ones, could have generated in sighted people unconscious connections and interpretations related to the aesthetics of the representation (although this was deliberately poor). The sample of 14 BVI users includes both congenital and gone blind people.

- **Scientific literacy**
  9 users are sighted Physics students, involved in the experimentation to pinpoint possible preconceptions that could influence the exploration and understanding of the proposed multi-sensory representation.

The final exhibit was shown to all these users, to verify the validity of the choices that were made during the process.

The sample of 22 users interviewed as part of the final evaluation of the exhibit was selected again based on the age, sensory impairment, and scientific literacy parameters. In this case, the two professional astronomers had experienced multi-sensory representations already in the past, while the other users did not. The main characteristics of this sample are:

- **Age**
  All selected users are adults.

- **Sight**
  We have involved both sighted (10) and BVI users (12), in order to pinpoint possible differences in the fruition and understanding as reported above. Also this sample of 12 BVI users includes both congenital and gone blind people.

- **Scientific literacy**
  2 professional BVI astronomers were included in the sample, in order to collect feedback from users proficient both in the scientific contents of the representation and in the fruition of non-visual representations. The scientific literacy of the sample of 20 visitors of the exhibition was the average one of the general public.

## 3.3  Represented data

The data used in all representations are the position, the intensity of the radio waves emission, the distance (if known), and the classification of astronomical sources emitting radio electromagnetic waves, achieved from several astronomical surveys. The position and intensity of the sources are taken from the radio astronomical surveys NRAO/VLA Sky Survey (NVSS[11], Condon et al. 1998) and Faint Images of the Radio Sky at Twenty-cm survey (FIRST[12], Becker et al. 1995), both taken with the Very Large Array (VLA) radio telescope in Socorro, New Mexico at 1.4GHz. The data about the distance and the source classification are taken from the optical survey VIMOS Public Extragalactic Redshift Survey (VIPERS[13], Guzzo et al. 2014) and Sloan Digital Sky Survey (SDSS[14], Blanton et al. 2017).

---

[11] NRAO/VLA Sky Survey: link
[12] FIRST survey: link

We chose different objects from various astronomical surveys, depending on the available data that we wanted to include into our representations. For the haptic-only representations, we used data from the survey NVSS, selecting both a close up on a single source named 0206+35 (with celestial coordinates R.A. = 02h 09m 38.9s and Dec = 35° 47' 5") and a wide field of 2 square degrees centered on the same source, showing 14 objects when selecting a minimum intensity of 4mJy/beam. For the multi-sensory representation, we selected a $18 \times 18$ arcmin region from the survey FIRST, centered at celestial coordinates R.A. = 2h 10m 19.992s and Dec = -4° 21' 27.07" showing six sources with intensity $> 2.2$ mJy/beam and distance data reported in the survey VIPERS. These parameters were selected to obtain a number of sources that presented different features (i.e., sources with different intensities, at different known distances and of different types). For sources that were not present in the VIPERS and SDSS surveys we had no information about the distance and the source type. We represented the lack of this information with white noise or blind bolts, as reported in the following sections.

### 3.4 Assessment

We performed a qualitative assessment of the two mono-sensory representations and of the resulting multi-sensory one, in three different steps. The goals of this first evaluation phase are:

- to assess the intelligibility of the sensory stimuli;
- to assess the autonomous understanding of the code;
- to assess the intelligibility of the meanings of the sensory stimuli, once the code has been made explicit;
- to identify possible misleading features of the representation.

To investigate such aspects, we divided each evaluation in two steps: (i) a first free and autonomous exploration of the representation, in which the only piece of information shared was: "The object of this representation is a celestial map" and (ii) a second exploration after the declaration of the code of representation, namely the mapping between the stimuli and the physical characteristics. Along with the declaration of the code, the astronomical concepts necessary to understand the representation, depending on the features represented in each case, were introduced.

For all representations, we explained what radio waves are, we compared them with optical light, and we discussed why they are not visible to human eyes. For the haptic-only representation, we introduced how two-dimensional maps represent a region of the celestial sphere. For acoustic-only representation, we discussed the distance from Earth of different celestial objects. For the multi-sensory representation, the meaning of intrinsic and apparent brightness of a source with respect to its distance from Earth was introduced, pointing out how some objects are intrinsically much brighter than others (thus represented by a larger number of bolts) even though further away (sounding at higher pitches). In this framework, also the difficulty of measuring the distance of objects in the sky was mentioned, to explain why two of the six represented objects did not have a recognizable sound associated (but only white noise). Finally, for the exhibit "Sense the Universe", we explained the main differences between galaxies and AGNs.

For the standalone version of the exhibit, we provided a written and audio description of the exhibit, the latter playing when touching a bolt attached to the explanatory panel. This was not included in the test with the 2 BVI astronomers and at the exhibition at "Palazzo delle Esposizioni", where the declaration of the code was left to the person guiding the experimentation or the fruition.

---

[13] VIPERS survey: link  
[14] SDSS survey: link

**Table 2.** Summary of the qualitative evaluation of the haptic, acoustic, and multi-sensory representations adopted in our study and their outcomes. The testing group consisted in 27 users, 13 Sighted and 14 BVI.

| Sensory representation | Adopted mapping | Outcomes and recommendations |
| --- | --- | --- |
| Haptic only | Relief points representing point-like sources. Relief denser and sparser textures representing regions at different intensities in close-up images of a source and contours to delimit them. | Relief points were easily associated with signal. Contours and diverse densities of points were hardly recognizable without an explicit code. |
| Acoustic only | Timbre-less sounds associated to point-like sources. Stereo sounds representing the 2D position of the sources in a sky region. Frequency of sound representing the frequency of the emitted electromagnetic wave (downscaled to fall in the earable domain). Different sound intensities (volume) representing different distances (same intensity of all sources) | The difference in frequency was hardly quantifiable (it is preferable to use sound with an identifiable timber). The frequency was sometimes associated with distances (better avoiding whatever connection to physical reality). The spatial position was mistaken with a 3D positioning (better avoiding stereo sounds for objects at astronomical distances). The differences in volume were hardly quantifiable (better using other parameters to represent distances). |
| Multi-sensory | An horizontal plywood tablet with columns of bolts representing point-like sources in a sky region. The position is represented both by physical position and by the stereo sound. Number of bolts in each column representing the source intensity. Frequency of the associated sound representing the distance. White noise representing unknown distances. | The representation was generally understandable. The decodification was straightforward. No connection with previous experience was reported. White noise was generally recognized as the lack of information about the represented feature (distance). Critical issues: 1) sound deactivates only by a second touch; 2) the tactile exploration is complicated by the bond of the electrical ground of the electronic board; 3) random bolt shapes are not representing a physical feature; 4) the physical distance of the point-like sources from the user (requiring a greater/smaller arm extension for the exploration) can be associated with the physical distance of the astronomical source from the observer. |

Our means of evaluations have been: 1) a questionnaire to be answered after the free exploration with these two open-ended questions: "What sensory stimuli do you perceive?" and "What information do you get from those sensory stimuli?"; 2) a questionnaire to be answered after the informed exploration, with the same questions as before plus an additional question: "Is there any information in contradiction with those gathered in your previous fruition?" 3) a final debrief during which we collected impressions and suggestions about the usability of the representation, discussed the real nature of the data and collected general feedback about the experience. Therefore the outputs were: two questionnaires for each participant and a report of the final debriefing. Since our goal was to collect as many hints as possible and thanks to a relative small number of participants, we performed direct text analysis. The results of this analysis guided the choices made and the adjustments implemented in each subsequent stage of the design.

The evaluation of the final exhibit was done through individual interviews with 2 BVI professional astronomers and with 20 (10 sighted and 10 BVI) of the many attendees of the "Inspiring Stars" exhibition at "Palazzo delle Esposizioni" in Rome, in January 2022.

# 4 IMPLEMENTATION AND PRELIMINARY TESTS

In the following we report the iterative design and evaluation process, and the assessment phase for the haptic, the acoustic and the first version of the multi-sensory representations. A summary of the experimentation process and outcomes of the three phases is reported in Table 2.

## 4.1 Design of haptic representation and preliminary tests

To design the mono-sensory haptic representation, we asked for the advice of professionals working at the Institute for the blinds Francesco Cavazza in Bologna (Italy) and in particular Dr. Loretta Secchi, curator of the Tactile Museum of Ancient and Modern Painting Anteros of the Institute[15]. Such Institutions are committed in research studies about the psychology of optical and tactile perception, also carrying on a pedagogical program for the design and creation of tactile reproductions of art works for the Museum.
We proposed to users two different haptic representations: one showing a region of the sky with different point-like sources and one representing regions at different intensities of a single object (Figure 1). We represented the point-like sources on the first map with relief points.

For the second representation, we represented the different intensities of the radio emission in different regions of the source with different spatial densities of the relief dots: denser (sparser) dots represent higher (smaller) intensities. In addition, we used contours to mark regions with different intensity of radio emission, to ease the interpretation of the transition in the signal intensity. Alternative options to represent the varying intensity of the source in the zoom-in map were considered and discarded, due to expert's advice: (i) the use of different volumes, namely dimensions or heights on the plane of the sheet, would give the idea of a "materiality" of the represented parameter, which instead is only energy; (ii) the use of materials with different roughness or textures of different shapes (triangles, squares, etc.) would have meant a limited number of variations, a difficulty in perceiving the transitions in small areas and a great memory effort in associating different materials with the corresponding intensities.

To create the haptic representation, we started from black and white images (Fig. 1), rendered in tactile form with the Minolta technique (Horsfall, 1997). This technique consists in drawing (or copying) the features to be relief-printed on a specific sheet of paper (available in A3 and A4 format) on which a layer of heat-sensitive micro-capsules is deposited. The sheet is then slid into an infrared oven. The heat causes the micro-capsules to swell, but only in the black-inked regions. With this technique we made two relief printings, representing the observed patch of the sky and the zoom-in on a specific radio source (Fig. 1, panels c and d).

---

[15] Istituto Cavazza: link

We tested these tactile maps with our sample. The presence of relief dots in the general map (Fig. 1, panel c) was easily associated with the presence of celestial objects. Instead, the contours of the zoom-in map of the single source (Fig. 1, panel d) were more difficult to perceive. For an effective and autonomous fruition of the representation, we chose to represent the general map, where each astronomical object was represented as one relief point only.

## 4.2 Design of acoustic representation and preliminary tests

We created the acoustic representation of data by using the software *audioscopio*[16]. To test the perception of the acoustic stimuli, we used a simple artificial set of data: 4 point-like sources, simulating 4 astronomical sources of radio waves all along the same axis (Fig. 2), each represented by a stereo signal. The stereo location of the sound corresponds to the actual location of the source on the x-axis. We chose to represent sources with the same intensity, which perceived volume depended on the distance from the observer. Different pitches correspond to the frequency of the emission of the source. For this evaluation we used fictitious intrinsic frequencies included in the earable domain. To represent frequencies of real radio data (roughly 3 kHz - 300 GHz), these would need to be downscaled. We have included two sources with the same intrinsic frequency (thus the same pitch of mapping sound), in order to assess if this feature of sound was easily perceivable, even if coming from different distances.

Sounds corresponding to the sources are synthetic sounds, all with the same timbre, and can be played by pressing on the computer keyboard number (1 to 4) associated with the source. They are heard in stereo through headphones. By pressing again the same number key, the corresponding source is turned off. All sounds can be heard simultaneously.

We tested the acoustic representation with our sample of users. As usual, we first let them freely explore the acoustic representation and then we declared the code. We collected feedback about the users perception and understandings both before and after the code declaration, using our questionnaires and a final discussion and debrief.

All users understood that each acoustic stimulus represented a single source. The spatial position of the sound was in all cases correctly interpreted as the spatial position of the emitting sources. However the users thought the sources were distributed in a 3D space, instead of the intended 2D one, namely on an imaginary 2D map "in front" of the user. After we declared the code, all the participants started to investigate the features of sound in order to link them with the physical properties of the radio emitting sources. All users but one could correctly interpret the spatial location of the sources. All the participants recognized the different sound intensities associating them to the distance of the source, and assuming that the intensity of the sources was the same. Nevertheless, the sound loudness was a hardly quantifiable feature, difficult to associate to actual differences in the distance. In addition, this result highlighted that intensity and distance should be represented with two different features, since the different volume could both mean objects with the same intensity at different distances and object with different intensities at the same distance or, even more confusing, a difference in the two variables at once, impossible to recognize from the variation of a single representative parameter. All participants but three perceived the different pitches associated to sounds, but showed difficulties in distinguishing higher from lower pitches, while none of them could tell when the difference in pitch between two sources was bigger or smaller. We conclude that possibly using the timbre of musical instruments instead of synthetic sounds would have improved the recognition of pitch changes. In addition to that, two Physics students (sighted) interpreted different pitches in terms of Doppler effect. This can be referred to the specific scientific literacy of the users, leading to a misinterpretation of the code (none of the sources was presented as moving) but is still worth attention.

---

[16] Github: link

During the debriefing, all participants reported having been puzzled during their first autonomous exploration, due to the fact that they were presented with only few stimuli, hardly conceivable as astronomical objects.

## 4.3 Design of the multi-sensory representation and preliminary tests

Building on the results of the tactile only and auditory only experimentations described in the previous sections, we created a multi-sensory representation. We represented point sources with relief objects on a 2D map. To create a correspondence between the sound and a tactile object we used a tool called *Makey Makey*®[17], designed by students of the Massachusetts Institute of Technology. This is an electronic board that can be connected to a computer through a USB cable. Crocodile clips are used to connect the Makey Makey® to the object that will have the function of a "key", namely the object that will activate the sound when the user touches it. The Makey Makey® board always needs a connection to a computer hosting the software that rules the conversion between the haptic input and the output sound. Using this board we created a plywood tablet where we inserted screws and bolts connected with the Makey Makey®. By touching one of these tactile objects, the user activates and deactivates the sound corresponding to that source. The spatial position of the tactile objects on the tablet corresponds to the position of the sound produced by the software *audioscopio*. In this way the position of the sources is linked to three sensory stimuli: the stereo position in the headphones and the visual and tactile position on the tablet. The shape and size of the screws and bolts on the tablet only depended on the availability on the market and the assembly requirements. The number of bolts corresponds to the intensity of the radio signal. We used sounds with the same loudness for all sources, as recommended by the results of the auditory-only experimentation. The frequency of the sound corresponds to the distance of the astronomical source from the observer (i.e. higher pitches representing object closer to the observer). To summarize: the representation consists in an horizontal tablet (2D map) in which objects are represented by a columns of bolts; the position of the column on the tablet and the stereo position of the sound represent the spatial position of the source in the sky; the number of bolts in each column represents the intensity of the source; the sound frequency represents the distance of the source.

The feedback collected from questionnaires showed that the stimuli were easily identified during the autonomous exploration and that the representation was clearly understandable after the code declaration. Some critical aspects were pointed out during the debriefing: 1) the fact that the users had to touch the objects not only to activate the sound, but also to deactivate it, complicated the exploration of the maps, especially when the users were using both hands, and generated confusion on the association sound-object; 2) the bond of having at least one finger on the electrical ground of the Makey Makey® complicated the tactile exploration of the map; 3) a different shape of the bolts at the top of the column could have been used as a representative parameter, since the users already expected this to be so; 4) one of the sighted users argued that the different distance from the user of the point-like sources on the map (implying a greater/smaller arm extension) could have been wrongly associated with the physical distance of the astronomical source from the observer (which was instead represented by the pitch of the signal). On the other side, BVI users didn't report the same issue, and when asked about it during the debriefing, they all answered that the positioning was not a problem for them, perhaps due to their better acquaintance with haptic exploration of 2D maps. Nonetheless, they suggested that an oblique positioning would have eased the exploration.

---

[17] Makey Makey: link

As for point 1), this was fixed by modifying the settings of the software *audioscopio*.

To solve point 2) instead we had to change board and decided to use the Bare Conductive® touch board[18] instead of the Makey Makey®, since the Bare Conductive® does not have an electrical ground. In addition to that, it can host a micro SD card with recorded sounds, without the need of a computer. This also made it easier to use and carry the exhibit, once the object-sound correspondence was set and allowed using sounds with a specific timber. We then abandoned the use of the software *audioscopio*, more adaptable for more complex representations, but not needed at this stage of the design. This meant losing the information about the two channels (stereo) acoustic stimulus, which anyway was considered as not necessary for representing objects at astronomical distances, for which the parallax (different apparent position from left or right ear) is not significant.

As for point 3), we decided to use the shape of the bolt at the top of the column to represent the type of represented source. This led us to browse the radio surveys for a region of sky in which objects of different type were present (galaxies, active galactic nuclei, ...). In the final representation we used an 18x18 arcmin observation centred at RA = 02h 10m 20.1124s and DEC = -04° 21' 27.079", from the survey FIRST. Optical data for the objects in the map are taken from the SDSS e VIPERS surveys (see Section "Represented data"). The region includes 3 elliptical galaxies, one active galactic nucleus and two objects without optical counterparts. The shapes of the top bolts chosen to represent different object were: convex bolts for galaxies, wing bolts for active galactic nuclei, blind bolts for unknown sources (Figure 3). We adopted white noise to represent the unknown distances, as this is generally recognizable as the lack of information about the represented feature, as confirmed by our experiment.

Finally, regarding point 4), we changed the positioning of the tablet from horizontal to oblique, also in order to ease the exploration.

## 5 RESULTS

On the basis of the findings of the preliminary test phases, we designed the final version of our multi-sensory representation. This was made of an oblique plywood tablet with columns of bolts representing six point-like sources in a sky region. The number of bolts in each column represented the source intensity. The shape of the top bolt represented the type of object. The sound associated to each source is a piano note, the pitch of which represents the distance of the source.
This representation became the exhibit "Sense the Universe", which entered the exhibition "Inspiring Stars" of the International Astronomical Union (link). It is shown in Figure 3 and was presented in several events, the last of which is the "Punti di vista" Festival held in January 2022 at "Palazzo delle esposizioni" in Rome.

We tested the the final exhibit "Sense the Universe" with an independent sample of 20 visitors (10 sighted and 10 BVI) of the exhibit and with two professional astronomers who are already familiar with sonification and haptic rendering of astronomical data, both for the professional community and the general public (see also Section 3.2). We conducted interviews with all the users.

All users referred not to have found any familiar feature in the representation and reported the intelligibility of the code after its declaration, suggesting that the representation was not misleading and that the concepts introduced in the declaration of the code were easily accessible, regardless previous knowledge and experience. The two BVI professional astronomers, who had both a high scientific literacy and some previous experience with multi-sensory representations, suggested the inclusion of additional stimuli in the representation.

---

[18] Bare Conductive touch board starter kit: link

Here we report the outcomes of this evaluation.

- **Intelligibility of the sensory stimuli**: all stimuli were easily perceived by all users.
- **Autonomous intelligibility of the code**: each varying stimulus was associated with a varying parameter, but only the correspondence between position on the map and position in the region of the sky was correctly argued by all; the correspondence between the number of bolts and the intensity of the source was supposed by 2 sighted and 4 BVI users (among which the two professional astronomers);
- **Intelligibility after the code declaration**: in all cases the code of representation was easily understood and applied after its declaration, and found very simple and straightforward. The two professional astronomers suggested to increase the density of the represented objects, still carefully avoiding confusion, and add other sensory stimuli (such as vibrations, pulsating sounds or lights) to represent additional features of the objects or to pair and reinforce other sensory stimuli (*redundancy*) and to make the representation also usable and understandable by deaf users. They also commented that adding the indication of the measurement units of the represented intensity (i.e. the value of a single bolt) would help the interpretation.
- **Possible misleading features**. None reported discrepancies between the stages before and after the code declaration. We also noticed that none of the users could autonomously interpret the representation before the declaration of the code. This highlights the importance of providing written and/or audio descriptions of multi-sensory representations to make the code explicit. We were able to better analyse this issue thanks to the debriefings and the interviews, as discussed in the following.

Here we report some additional issues that have come up both during the debriefing with the testing users of the main experimentation and with the general public that have tried the exhibit "Sense the Universe" in a museum context.

**The analogy with experience**

Both the preliminary multi-sensory and the final exhibit proven to be effective for one of our main aims, i.e. creating a meaningful experience without leading to misunderstandings and misinterpretation due to an uncontrolled connection with reality and previous experience. All the 22 interviewed users of the final exhibit reported that they could not find any "analogy with experience", meaning they did not recognize any of the represented features as familiar. We consider this a successful result of the design process, where we iteratively excluded a number of features that the users connected with previous experiences and could lead to misconceptions about the physics of the represented objects.

Not finding any evident connection with their personal experience of representations of the Universe, many users expressed a certain sense of awkwardness. During the free experimentation phase, before the code declaration, all the users involved in the experimentation found it hard to realize what the relevant elements of the representation were and what was their meaning ("I don't know what to look for" was a common comment). This, while implying an initial discomfort, guaranteed deep attention and promptness with regards to the representation.

**Emotional involvement**

There was very little or no emotional commitment in the exploration phase during the experimentation. We noticed a little more involvement in users experiencing the exhibit "Sense the Universe" in the museum exhibition (which may be due to the more engaging environment and the related different attitude. In both cases though, we observed how this "emotional detachment" made it easier to recognize and accept the arbitrariness of the presented code and to correctly interpret the data. However some users, especially the BVI involved in the experimentation, reported a little disappointment, as they expected a more engaging experience.

During the debriefing that followed the experimentation, the emotional aspect was restored and the reaction was generally very positive. We noticed a very positive response from BVI users, when it became clear to them that none can perceive the represented phenomena with any sense, as they are invisible, not touchable, and impossible to hear. One of the BVI users said, excited: "ah, then we are even!", showing how multi-sensory representations can be key to engage such users with astronomy and scientific disciplines, using a thoughtful emotional involvement driven by the object of the representation instead of the one inspired by the aesthetics of the representation (the so-called "wow" effect).

## 6 DISCUSSION AND CONCLUSIONS

In this study we argue how multi-sensory representations of scientific, and in particular astronomical, data are important both in outreach and educational contexts to offer users and learners a fair accessibility to culture as well as an effective understanding, free of misconceptions. We created an exhibit, "Sense the Universe", representing astronomical sources such as galaxies and active galactic nuclei, and their characteristics (e.g., intensity, position in the sky, distance, source type) taken from radio astronomical surveys. These data are not directly visible as our sight is not sensitive to radio wavelengths. We made an arbitrary multi-sensory representation involving the sense of sight, hearing, and touch. First we designed and tested each mono-sensory representation separately and then, on the basis of such experimentation results, we created and evaluated the multi-sensory exhibit.

The design of arbitrary (non-figurative), multi-sensory, multi-modal and equitable (universal) representations requires a very attentive and iterative process of thinking-testing-evaluating. Each phase of the iteration should involve experts, self-advocates and final users, in order to assess the usability and effectiveness of the representation, and also to pinpoint possible misleading features.

Our experimentation was carried out with constant advice from experts and through an iteration process with the users/learners, in analogy with the design-based learning methodology, in which the users are engaged as co-designer of the artifact that they will use for an informal learning experience.

The exhibit "Sense the Universe" offers multiple means of engagement, representation and action (as for the guidelines of Universal Design for Learning). The use of multi-sensory stimuli was studied with regards to its suitability in a proper meaning-making process. In Table 2 we outline some methodological guidelines to be used by future studies representing astronomical data with multi-sensory stimuli, in order to maximize their intelligibility and effectiveness. In particular, we argue that it is important to establish non-ambiguous connections between the characteristics of the sensory stimulus and the represented features, to avoid misunderstandings and unforeseen connections with memory and experience.

Though being arbitrary, our representations led to at least two analogies to previous experiences, one related to the scientific literacy of the Physics students (two of which described the different sound frequency associated to the distances of the sources in terms of Doppler effect) and the other perhaps related to the lack of practice in haptic exploration (one sighted user referred that objects that required greater extension of the arms to be touched looked as more distant). We conclude that, even for sensory representations other than visual ones, great attention must be paid to concepts that may inspire analogies with any existing knowledge or personal experience. In general, we argue that it is key to always accompany representations (both figurative and arbitrary ones) with the explicit definition of the code of representation or at least the declaration of their disconnection from reality.

The use of arbitrary representations implies a certain degree of "emotional detachment", due to the poor and awkward aesthetics of the representation, and/or to the training needed to understand and memorize the code, with the effect of slowing down the exploration process. This can actually lead to a more thoughtful fruition and to a deeper understanding of the represented reality, calling into action different modalities of interaction and competences (musical, logical, spatial, and bodily-kinesthetic, linguistic). In addition to that, we believe the initial discomfort in looking for relevant elements that the users reported is particularly well-fitting and appropriate in the context of the represented data, as it shows the method used by researchers to proceed in the exploration of astronomical data. At the beginning everything seems equally relevant and the researcher has to decide what to focus on, also basing on the perception of different features of the data.

We conclude that multi-sensory representations of astronomical data are a promising way to convey in an universal and effective way scientific and in particular astronomical concepts to the general public and the students in schools.

## 6.1 Future work

We encourage future studies to further test our results and possibly validate or update the guidelines that we highlight in Table 2. In particular, it will be important to evaluate our recommendations (and the exhibit "Sense the Universe") with larger samples of users. It will also be interesting to further study whether such recommendations depend on the scientific literacy of the audience, on the level of their sensory impairment, and/or on their cultural background.

A natural extension of the exhibit "Sense the Universe" is the design of a multi-sensory planetarium dome where celestial objects are represented as visual, tactile and acoustic stimuli. The design and pilot testing of such an exhibit is currently ongoing.
Furthermore, we are designing the educational activity "Make the Universe" based on the experience obtained with this study. The activity will consist in the discussion, group design and "do it yourself" production of multi-sensory displays of the Universe, that make use of visual, tactile and sound materials. All these projects and products are going to be tested both in museums and schools with a dedicated evaluation protocol.

## CONFLICT OF INTEREST STATEMENT

The authors declare that the research was conducted in the absence of any commercial or financial relationships that could be construed as a potential conflict of interest.

## AUTHOR CONTRIBUTIONS

SV designed the study performed the evaluation and user testing. SV and AZ discussed the results and wrote the paper.

**FUNDING**

The authors declare that no fundings were involved in supporting this work.

**ACKNOWLEDGMENTS**

The authors thank Dr. Fernando Torrente and Dr. Loretta Secchi of the Istituto Francesco Cavazza (Bologna, Italy), Prof. Barbara Pecori (tutor of the PhD study leading to this project), Francesco Bedosti (author of the software audioscopio) and the astronomers Nicholas Bonne and Enrique Pérez Montero for their support and valuable discussions during the development of this project.

**DATA AVAILABILITY STATEMENT**

The datasets analyzed for this study can be made available upon reasonable request to the authors.

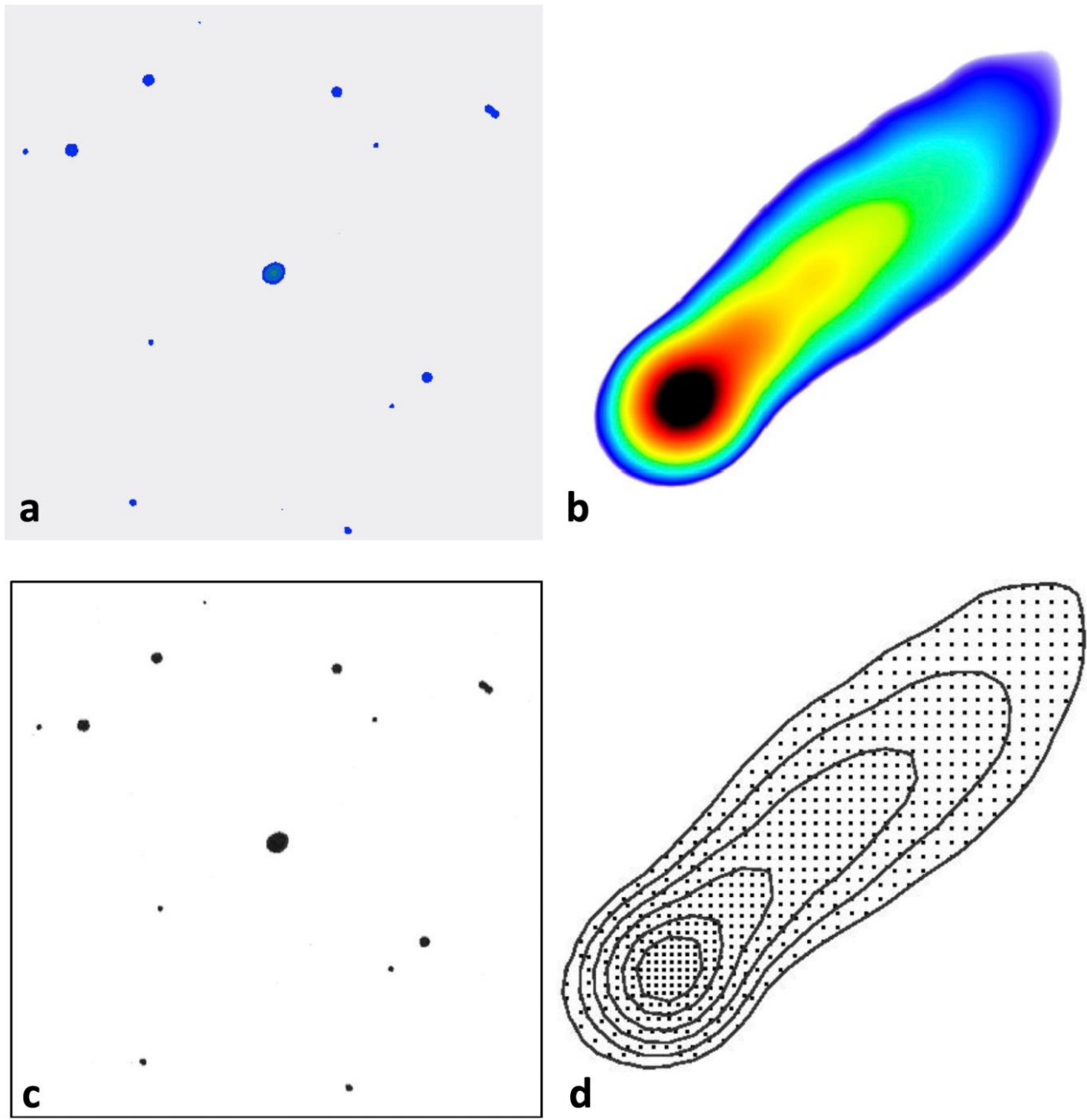

**Figure 1.** Haptic representation of the radio sources. **Panel a**: false-color image of the radio waves emitted at 1.4GHz (data NRAO/VLA Sky Survey, NVSS) by a 2 square degrees sky region centered on the source called 0206+35 (Right Ascension 02h 09m 38.9s and Declination 35° 47' 57"). Blue color represents regions at intensity higher than 4 mY/beam. **Panel b**: false color image of the radio emission of source 0206+35 at 1.4GHz (data NVSS). Colors blue, green, yellow, orange and red represent the ranges of intensity between the contour levels reported in Panel d. Black color represents the region with intensity from ~ 62 mJy/beam to the maximum value of ~ 1.3 Jy/beam, with a beam of 1,7arcseconds for VLA images at 1,4 GHz. **Panel c**: visual rendering of the tactile map used to represent the sky region centered on the source 0206+35. **Panel d**: Intensity contours of the source 0206+35. at levels of 0.00827; 0.0136; 0.021; 0.032; 0.047 and 0.062 mJy/beam. The points inside each contour thin out from the center toward the outskirts, as the intensity of the radio waves emission.

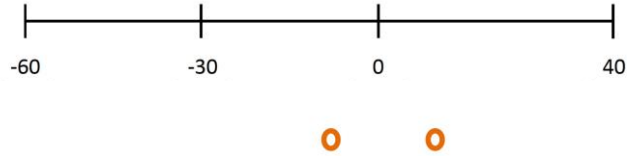

**Figure 2.** Scheme of the spatial position of sources for the auditory representation of data. Numbers indicate the x-position (in centimeters) of the four represented sources on a horizontal axis centered in front of the user. The position of the ears (represented by the 2 red circles) is -10 cm for the left ear and +10 cm for the right one, and corresponds to the position of the headphones in which each source sound as a stereo signal. The stereo location of the sound corresponds to the actual location of the source on the x-axis.

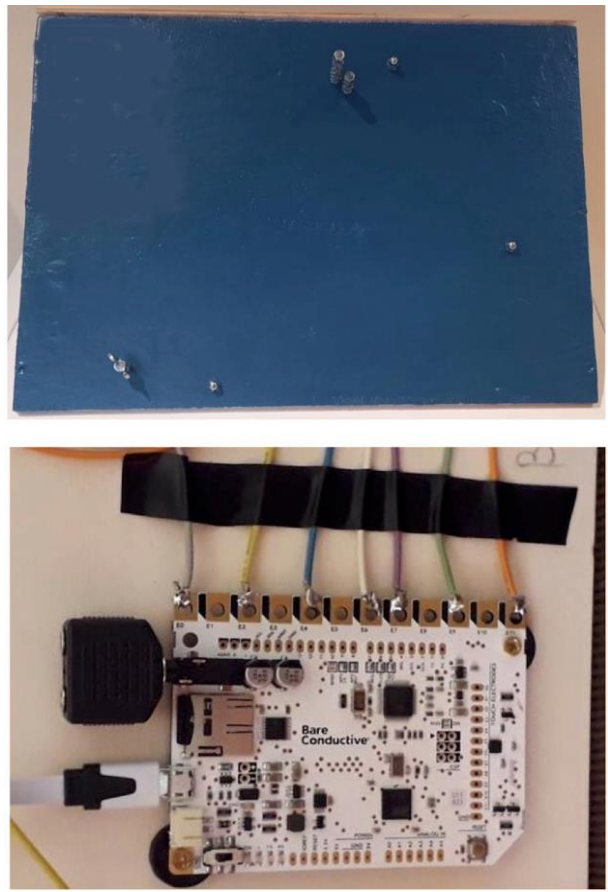

**Figure 3.** The multi-sensory exhibit "Sense the Universe". **Top panel:** front of the tablet, showing the bolts used to represent sources. **Bottom panel:** electronic board used to reproduce sounds.